\documentclass[11pt]{article}

\usepackage[margin=1in]{geometry}
\usepackage{amsmath,amssymb,amsthm,mathtools}
\usepackage{booktabs}
\usepackage{graphicx}
\usepackage[nomarkers,nolists,tablesfirst]{endfloat}
\usepackage{natbib}
\usepackage{microtype}
\usepackage[hidelinks]{hyperref}

\newtheorem{theorem}{Theorem}

\newcommand{\Rset}{\mathcal R}
\newcommand{\Rzero}{\mathcal R_0}

\newcommand{\thetamin}{\theta_{\min}}

\newcommand{\thetahatmin}{\widehat\theta_{\min}}

\title{Worst-Case Win Ratios Under Partially Specified Outcome Hierarchies}

\author{
Kexuan Li\textsuperscript{1},
Xue Fan\textsuperscript{2},
Lingli Yang\textsuperscript{3}
\\[0.8em]
\small \textsuperscript{1}Department of Biostatistics, Bristol Myers Squibb, USA\\
\small \textsuperscript{2}Department of Biostatistics and Data Science, University of Texas Houston, USA\\
\small \textsuperscript{3}Takeda Pharmaceuticals, USA
}

\date{}

\begin{document}
\maketitle

\begin{abstract}
Win statistics require a prespecified outcome hierarchy. Clinical teams sometimes agree only on the highest priority outcome, leaving the order of lower priority outcomes unresolved, and clinically meaningful thresholds may be specified as ranges. Separate sensitivity analyses describe how the results change. A single inference for the full set of planned analyses is generally absent. We define the estimand as the smallest net benefit among the comparison rules allowed by the protocol or statistical analysis plan. Each rule uses the usual two-sample U-statistic. Large-sample results are developed for a finite list of rules and for a continuous threshold range. Inversion of an intersection-union test gives a one-sided lower confidence bound, with no multiplicity adjustment for the single claim that every individual net benefit is positive. Simulations show valid one-sided coverage and illustrate the gap between a favorable result for one selected hierarchy and a favorable result across all prespecified hierarchies. An application to ACTG 175 shows how the ordering of laboratory outcomes can affect the strength of the conclusion. A further example identifies an unfavorable threshold that is missed by a sparse grid.
\end{abstract}


\section{Introduction}
Composite endpoints are common in clinical trials when no single outcome adequately describes the clinical experience of a participant. A conventional time-to-first-event analysis treats all component events as equally decisive at the time of their first occurrence. This can be unsatisfactory when the components differ substantially in clinical importance. The Finkelstein--Schoenfeld procedure and generalized pairwise comparisons address this problem by comparing participants across clinically ordered outcomes \citep{finkelstein1999,buyse2010}. The win ratio gave this construction a direct effect measure and helped establish its use in cardiovascular and other trials \citep{pocock2012}. Related measures include the net benefit and win odds. They are based on the same pairwise win and loss probabilities and test closely related null hypotheses \citep{dong2023,verbeeck2023}. For a fixed hierarchy, the win statistic is a rank statistic: every treated participant is compared with every control participant and the resulting estimator is a two-sample U-statistic. This observation underlies variance estimators, confidence intervals, regression methods, and extensions for censoring and recurrent events \citep{luo2015,bebu2016,oakes2016,mao2019,mao2024estimand}. 

In the literature, most of the developments of win statistics take the comparison rule as fixed, however, in practice, a trial team may nevertheless agree that death belongs first while leaving two nonfatal outcomes unordered. The clinically meaningful difference for a component may also be specified as a range. Lower tiers and threshold choices can materially affect the analysis \citep{shoji2025}, and the clinical basis for a hierarchy is often poorly documented \citep{ehrenzeller2026}. To overcome this chanllange, multiple approaches have been proposed. To name a few, threshold sensitivity analyses describe results over an inspected grid \citep{ozenne2025}. Multiple alternating thresholds place several thresholds within one comparison procedure \citep{mou2026thresholds}; rotation win statistics average hierarchies when selected outcomes have equal priority \citep{mou2025rotation}; and the win ratio for partially ordered data permits incomparable response pairs under a fixed partial order \citep{mao2026poset}. However, the methods mentioned above may still be lacking in clinical interpretation. For example, net benefits of 0.08 and $-0.02$ under two clinically reasonable hierarchies have a positive average, although one hierarchy favors control. To provide a more clinical meaningful interpretion, we propose to consider the smallest net benefit across all hierarchies and thresholds judged clinically reasonable before treatment assignments are examined. We give inference for both a finite list and a continuous threshold range using familiar rank-statistic arguments.
\section{Clinical comparison rules and estimand}

Consider a two-arm randomized trial. Let $X$ contain the endpoint components for a randomly selected participant assigned to treatment, and let $Y$ contain the same components for an independently selected control participant. Typical components include death status, number of hospitalizations, and a quality-of-life score through a common follow-up time. For outcome $k$, the treated and control participants are compared by
\begin{equation}
c_k(x_k,y_k;\tau_k)
=
I\{s_k(x_k-y_k)\geq \tau_k\}
-I\{s_k(y_k-x_k)\geq \tau_k\},
\label{eq:component-score}
\end{equation}
where $s_k=1$ if a larger value is better, $s_k=-1$ if a smaller value is better, and $\tau_k\geq0$ is the clinically meaningful difference for that outcome. Thus $c_k=1$ is a win for treatment, $c_k=-1$ is a loss, and $c_k=0$ means that the difference does not reach the threshold and the comparison moves to the next outcome. Binary and ordinal outcomes use the same win, loss, and tie coding with the appropriate outcome-specific comparison.

Let $\pi=(\pi_1,\ldots,\pi_K)$ give the order in which the $K$ outcomes are compared, and let $\tau=(\tau_1,\ldots,\tau_K)$ contain their thresholds. One complete comparison rule is $r=(\pi,\tau)$. Its pairwise win-loss score is
\begin{equation}
h_{(\pi,\tau)}(x,y)
=\sum_{m=1}^{K}
c_{\pi_m}(x_{\pi_m},y_{\pi_m};\tau_{\pi_m})
\prod_{\ell<m}
I\{c_{\pi_\ell}(x_{\pi_\ell},y_{\pi_\ell};\tau_{\pi_\ell})=0\}.
\label{eq:lex-score}
\end{equation}
We write $h_r$ as shorthand for $h_{(\pi,\tau)}$. On the right-hand side of Equation~\eqref{eq:lex-score}, $\pi$ determines the hierarchy and $\tau$ determines the thresholds. The product carries the comparison forward after ties on all higher priority outcomes. With the hierarchy death, hospitalization, and quality of life, a tie on death leads to hospitalization; ties on both outcomes lead to quality of life. The final value is 1 for a treatment win, $-1$ for a treatment loss, and 0 for a tie.

The protocol or statistical analysis plan may leave more than one hierarchy clinically permissible. Death, for example, may be required to come before hospitalization while either an urgent visit or a patient-reported outcome may occupy the next position. Let $\Pi$ be the resulting list of hierarchies and let $\mathcal T$ contain the prespecified threshold values or ranges. The analysis set is
\begin{equation}
\Rset=\{r=(\pi,\tau):\pi\in\Pi,\ \tau\in\mathcal T\}.
\label{eq:rule-set}
\end{equation}
Here $r$ indexes one planned combination of a hierarchy and thresholds. Selected threshold values give a finite set $\Rset=\{r_1,\ldots,r_M\}$. If the plan includes every value in a clinically meaningful range, such as $\tau_Q\in[5,15]$, $\Rset$ includes the full interval. Specification takes place before examination of unblinded treatment differences so that the treatment effect remains prospectively defined.

For a rule $r$, define
\begin{align}
p_W(r)&=\Pr\{h_r(X,Y)=1\},
&p_L(r)&=\Pr\{h_r(X,Y)=-1\},\\
\theta_r&=p_W(r)-p_L(r),
&WR_r&=\frac{p_W(r)}{p_L(r)}.
\label{eq:rule-effects}
\end{align}
The parameter $\theta_r$ is the net benefit for one hierarchy and one set of thresholds. It is the proportion of cross-arm pairs that favor treatment minus the proportion that favor control. For example, $\theta_r=0.08$ means that, per 100 treated-control pairs, there are eight more treatment wins than treatment losses. If $p_L(r)>0$, then $\theta_r>0$ if and only if $WR_r>1$.

The treatment effect of interest is the smallest net benefit among all prespecified analyses:
\begin{equation}
\thetamin=\inf_{r\in\Rset}\theta_r.
\label{eq:minimum-effect}
\end{equation}
Once $\Rset$ has been prespecified, $\thetamin$ is a population estimand defined by the planned analyses. Its variation across rules is separate from uncertainty due to missing data. The condition $\thetamin>0$ means that treatment produces more wins than losses under every hierarchy and threshold in $\Rset$. Net benefit is used for the main theory because it is bounded and its standard error remains stable when the loss probability is small. The clinical report can also present win ratios for the individual analyses, including the one with the smallest estimated net benefit.

In practice, $\Rset$ is determined by the clinically acceptable choices in the protocol or statistical analysis plan. Suppose death is always compared first, hospitalization and quality of life may appear in either order, and the quality-of-life threshold may be 5, 10, or 15 points. These choices give six rules, and $\thetamin$ is the smallest net benefit among the six corresponding analyses. If every threshold between 5 and 15 points is considered acceptable, the minimum is taken over the entire interval. This estimand is useful when the trial conclusion is expected to remain favorable across all such choices. 

\section{Estimation and statistical inference}

Let $X_1,\ldots,X_{n_1}$ and $Y_1,\ldots,Y_{n_0}$ be independent samples from the treatment and control populations, and let $n=n_1+n_0$. For $r\in\Rset$, estimate the net benefit by
\begin{equation}
\widehat\theta_r
=\frac{1}{n_1n_0}
\sum_{i=1}^{n_1}\sum_{j=1}^{n_0}h_r(X_i,Y_j),
\qquad
\thetahatmin=\inf_{r\in\Rset}\widehat\theta_r.
\label{eq:estimator}
\end{equation}
For a given analysis rule, $\widehat\theta_r$ is the average of the $n_1n_0$ pairwise scores: a treatment win contributes 1, a loss contributes $-1$, and a tie contributes 0. The estimate $\thetahatmin$ is the smallest of these results. The individual estimates should also be reported because they show which hierarchy or threshold changes the clinical conclusion.

Each participant appears in many pairs, which induces dependence among the $n_1n_0$ comparisons. The usual U-statistic variance estimator handles this dependence through participant-level terms. For a finite set $\Rset=\{r_1,\ldots,r_M\}$, write $\widehat{\boldsymbol\theta}=(\widehat\theta_{r_1},\ldots,\widehat\theta_{r_M})^\mathsf T$ and define
\begin{align}
\widehat a_{ir}
&=\frac{1}{n_0}\sum_{j=1}^{n_0}h_r(X_i,Y_j)-\widehat\theta_r,\\
\widehat b_{jr}
&=\frac{1}{n_1}\sum_{i=1}^{n_1}h_r(X_i,Y_j)-\widehat\theta_r.
\label{eq:projections}
\end{align}
The term $\widehat a_{ir}$ summarizes how treated participant $i$ contributes to the estimated net benefit under rule $r$; $\widehat b_{jr}$ has the same interpretation for control participant $j$. Let $\widehat{\boldsymbol a}_i$ and $\widehat{\boldsymbol b}_j$ collect these quantities over the $M$ rules. The covariance estimator is
\begin{equation}
\widehat V
=\frac{1}{n_1(n_1-1)}\sum_{i=1}^{n_1}
\widehat{\boldsymbol a}_i\widehat{\boldsymbol a}_i^\mathsf T
+\frac{1}{n_0(n_0-1)}\sum_{j=1}^{n_0}
\widehat{\boldsymbol b}_j\widehat{\boldsymbol b}_j^\mathsf T.
\label{eq:covariance}
\end{equation}
The off-diagonal elements account for the fact that the same trial participants are analyzed under every hierarchy and threshold. Denote the standard error of $\widehat\theta_r$, obtained from the corresponding diagonal element, by $\widehat s_r$.

\subsection{Finite rule sets}

For $r\in\Rset$, define
\begin{align*}
a_r(x)&=E\{h_r(x,Y)\}-\theta_r,\\
b_r(y)&=E\{h_r(X,y)\}-\theta_r,
\end{align*}
and let $\boldsymbol a(x)$ and $\boldsymbol b(y)$ collect these functions.

\begin{theorem}[Joint large-sample distribution]
\label{thm:joint}
Suppose $n_1/n\to\lambda\in(0,1)$ and $\Rset$ is finite. Then
\begin{equation}
\sqrt n\{\widehat{\boldsymbol\theta}-\boldsymbol\theta\}
\ \rightsquigarrow\ 
\boldsymbol G\sim N_M(0,\Sigma),
\label{eq:joint-clt}
\end{equation}
where
\begin{equation}
\Sigma
=\lambda^{-1}\operatorname{Var}\{\boldsymbol a(X)\}
+(1-\lambda)^{-1}\operatorname{Var}\{\boldsymbol b(Y)\}.
\label{eq:sigma}
\end{equation}
Moreover, $n\widehat V\to_p\Sigma$.
\end{theorem}

Theorem~\ref{thm:joint} is the usual joint central limit theorem for the planned win-statistic analyses. The bounded win, loss, and tie score makes the moment conditions automatic. Estimating the full covariance is important because the same participants and many of the same pairwise decisions are used in every analysis.

Let
\begin{equation}
\mathcal R_{\min}=\arg\min_{r\in\Rset}\theta_r
\label{eq:minimum-rules}
\end{equation}
be the set of hierarchies and thresholds that attain the smallest true net benefit.

\begin{theorem}[Limit of the minimum net benefit]
\label{thm:minimum}
Under the conditions of Theorem~\ref{thm:joint},
\begin{equation}
\sqrt n(\thetahatmin-\thetamin)
\ \rightsquigarrow\
\min_{r\in\mathcal R_{\min}}G_r.
\label{eq:min-limit}
\end{equation}
If $\mathcal R_{\min}$ contains one rule, the limit is univariate normal. If it contains more than one rule, the limit is the minimum of correlated normal variables and is not generally normal.
\end{theorem}

When two planned analyses have the same smallest true effect, sampling variation can change which one has the smallest observed estimate. A conventional normal interval applied after selecting that analysis ignores this switching and may have incorrect coverage. Mathematically, the minimum is directionally differentiable at such points, although full differentiability fails when more than one rule attains the minimum, as in other problems involving several inequality constraints \citep{chernozhukov2013,fang2019}.

\subsection{Continuous threshold intervals}

Suppose that the number of hierarchies in $\Pi$ remains finite and that one or more thresholds vary over a closed, bounded range $\mathcal T$. Let $\mathcal H=\{h_r:r\in\Rset\}$ be the corresponding collection of pairwise score functions. The next result covers the full threshold range.

\begin{theorem}[Continuous threshold sets]
\label{thm:process}
Assume that $n_1/n\to\lambda\in(0,1)$, $\Rset$ is compact, $\theta_r$ is continuous in $r$, and $\mathcal H$ is a measurable bounded class for which the two-sample U-process is asymptotically equicontinuous. Then, in $\ell^\infty(\Rset)$,
\begin{equation}
\sqrt n\{\widehat\theta_r-\theta_r:r\in\Rset\}
\ \rightsquigarrow\
\{G(r):r\in\Rset\},
\label{eq:uprocess}
\end{equation}
where $G$ is a tight mean-zero Gaussian process with covariance determined by the treatment and control first-order projections. In addition,
\begin{equation}
\sup_{r\in\Rset}|\widehat\theta_r-\theta_r|\to_p0,
\qquad
\sqrt n(\thetahatmin-\thetamin)
\rightsquigarrow
\inf_{r\in\mathcal R_{\min}}G(r).
\label{eq:continuous-min-limit}
\end{equation}
\end{theorem}

Theorem~\ref{thm:process} gives uniform consistency of the estimated net-benefit curve over the prespecified threshold range, with inference based on the smallest value of that curve. For the uncensored score in Equation~\eqref{eq:lex-score}, its technical conditions hold when the number of hierarchies is finite, thresholds vary over closed and bounded intervals, the threshold comparison classes are of VC type, and the relevant pairwise differences have no probability mass exactly at a moving threshold. Indicator threshold classes are VC classes, and finite sums, products, and unions preserve the required properties \citep{arcones1993,vdvaartwellner1996}. Discrete outcomes can remain in the endpoint; continuity is required only for an outcome whose threshold varies over an interval.

The limit in Equation~\eqref{eq:continuous-min-limit} has the same interpretation as in the finite case. It is normal when a single hierarchy and threshold attain the smallest true net benefit. With several minimizers, it is the minimum of the corresponding Gaussian process. The minimum across analyses creates the nonstandard limit; each fixed-threshold estimator remains a regular U-statistic.

\subsection{Confidence bound for benefit under all rules}

The clinical claim of interest is that treatment has a positive net benefit under every prespecified hierarchy and threshold. The corresponding hypothesis is
\begin{equation}
H_0:\text{there is at least one }r\in\Rset\text{ with }\theta_r\leq0
\quad\text{versus}\quad
H_1:\theta_r>0\text{ for every }r\in\Rset.
\label{eq:iut-hypothesis}
\end{equation}
For each planned analysis, calculate the usual one-sided test statistic and p-value,
\begin{equation}
Z_r=\frac{\widehat\theta_r}{\widehat s_r},
\qquad
p_r=1-\Phi(Z_r).
\label{eq:rule-pvalues}
\end{equation}
The intersection-union p-value is
\begin{equation}
 p_{\mathrm{IUT}}=\sup_{r\in\Rset}p_r.
\label{eq:iut}
\end{equation}
Thus the overall p-value is the largest of the individual p-values, and the null is rejected when every planned analysis is significant at level $\alpha$. No conventional multiplicity adjustment is needed for this single claim. This is the standard intersection-union principle \citep{berger1982}. Under the null, at least one of the individual null hypotheses is true, so the probability of rejecting all of them is no larger than $\alpha$. Adding more hierarchies or thresholds can reduce power, but it also makes the clinical claim stronger.

Applying the same test to candidate values of $\thetamin$ and inverting it gives the one-sided lower confidence bound
\begin{equation}
L_{1-\alpha}
=\inf_{r\in\Rset}
\left\{\widehat\theta_r-z_{1-\alpha}\widehat s_r\right\},
\label{eq:lower-bound}
\end{equation}
where $z_{1-\alpha}$ is the standard normal quantile. In practice, the analyst calculates the usual lower confidence limit for every planned analysis and reports the smallest one. Retaining every rule-specific limit accounts for the data-dependent identity of the smallest point estimate.

\begin{theorem}[Validity of the lower confidence bound]
\label{thm:iut}
Suppose the conditions of Theorem~\ref{thm:joint} hold for a finite rule set, or the conditions of Theorem~\ref{thm:process} hold for a compact rule set. Assume that the standard error is consistent and positive at every population minimizer. Then
\begin{equation}
\liminf_{n\to\infty}P\{\thetamin\geq L_{1-\alpha}\}\geq1-\alpha.
\label{eq:coverage}
\end{equation}
The test that rejects $H_0:\thetamin\leq0$ when $L_{1-\alpha}>0$ has asymptotic level no greater than $\alpha$. In the finite-rule case, if $\thetamin=0$, exactly one rule has zero effect, and all remaining rules have positive effects bounded away from zero, its limiting rejection probability is $\alpha$. It can be conservative when several rules are on the boundary.
\end{theorem}

Implementation of Theorem~\ref{thm:iut} bypasses estimation of the hierarchies or thresholds that attain the smallest true effect. The bound can be conservative when several individual effects are equal or nearly equal, while retaining validity throughout the composite null.

Proofs of all four theorems are given in Supporting Information Appendix S1.

The same construction can be used for the win ratio when the loss probability is not close to zero: calculate a one-sided lower limit for each log win ratio, exponentiate, and report the smallest limit. We use net benefit for the main inference because its standard error is well behaved even when losses are uncommon.

For one continuous threshold, the observed net benefit changes only when the threshold passes an observed absolute difference between a treated and a control participant. Between two such values, the pairwise decisions, point estimate, and standard error remain unchanged. The minimum in Equation~\eqref{eq:lower-bound} can therefore be found by evaluating the interval boundaries and these observed values. With several continuous thresholds, the same idea applies but computation may become demanding.

For sample-size planning, Equation~\eqref{eq:joint-clt} gives the joint normal approximation to the individual test statistics. Power is the probability that every statistic exceeds $z_{1-\alpha}$ under the assumed outcome distributions. The planning assumptions must give $\theta_r>0$ for every $r\in\Rset$; a negative population effect for any prespecified rule precludes the planned claim, irrespective of sample size.

\section{Simulation study}

The simulations were designed to evaluate one-sided confidence coverage, type I error, power, and the ability to distinguish a favorable result under one selected analysis from a favorable result under every prespecified analysis. Each participant had three outcomes: death $D$, number of hospitalizations $H$, and a quality-of-life score $Q$. Smaller values were favorable for $D$ and $H$, and larger values were favorable for $Q$. Given treatment indicator $A$ and latent severity $S\sim N(0,1)$, data were generated from
\begin{align}
D&\sim\operatorname{Bernoulli}\left[\operatorname{expit}\{-2.2+\beta_D A+0.60S\}\right],\\
H&=\min\left[\operatorname{Poisson}\{\exp(-0.8+\beta_H A+0.35S)\},3\right],\\
Q&=\beta_QA-0.55S+\epsilon,
\qquad \epsilon\sim N(0,0.9^2).
\label{eq:dgp}
\end{align}
Death was always compared first. The remaining order was either $D$--$H$--$Q$ or $D$--$Q$--$H$, and the threshold for $Q$ was $0.25$, $0.75$, or $1.25$ standardized units, giving six prespecified analyses. Any difference in hospitalization count was considered decisive.

Table~\ref{tab:scenarios} lists the scenarios. In the scenario where the sign differed by threshold, the residual standard deviation for $Q$ was $0.65$ and the treatment shift was $-0.40$ with probability $0.85$ and $2.50$ with probability $0.15$. This mixture represents a setting in which many participants have a small unfavorable change while a minority have a large favorable change, so the sign can depend on the threshold. Population effects were approximated with five million independent treated-control pairs, except that the global-null effects are exactly zero by exchangeability.

\begin{table}[t]
\centering
\caption{Data-generating scenarios. A negative coefficient is beneficial for death and hospitalization; a positive coefficient is beneficial for quality of life.}
\label{tab:scenarios}
\resizebox{\textwidth}{!}{%
\begin{tabular}{lrrrl}
\toprule
Scenario & $\beta_D$ & $\beta_H$ & $\beta_Q$ & Purpose \\
\midrule
Global null & 0.00 & 0.00 & 0.00 & All six rules on the boundary \\
One active null rule & $-0.10$ & $-0.40$ & $-0.050$ & One rule slightly below zero; others favorable \\
Weak benefit under all rules & $-0.10$ & $-0.20$ & 0.12 & Modest benefit under all rules \\
Strong benefit under all rules & $-0.25$ & $-0.35$ & 0.35 & Clear benefit under all rules \\
Different signs by hierarchy & $-0.10$ & $-0.80$ & $-0.35$ & Hospitalization benefit and quality-of-life harm \\
Different signs by threshold & $-0.05$ & $-0.05$ & mixture & Small common harm and large uncommon benefit in $Q$ \\
\bottomrule
\end{tabular}
}
\end{table}

We used $n_1=n_0=150$ and $n_1=n_0=300$ and 5,000 simulation replicates. The one-sided significance level was $0.025$. We evaluated the lower confidence bound in Equation~\eqref{eq:lower-bound}, the corresponding intersection-union test, and a conventional analysis using only the hierarchy $D$--$H$--$Q$ with threshold $0.75$. The conventional analysis concerns that single hierarchy and threshold; the intersection-union test concerns the complete prespecified set.

The results are reported in Table~\ref{tab:simulation}. Coverage of the lower confidence bound for $\thetamin$ ranged from $0.974$ to $0.994$. Under the global null, rejection probabilities were $0.008$ and $0.006$, reflecting conservatism when all six individual null hypotheses were on the boundary. When only one rule was slightly below zero and the others were favorable, the rejection probabilities were $0.025$ and $0.023$, close to the one-sided level. Under weak benefit for all six rules, power increased from $0.097$ to $0.193$ as the sample size per arm increased from 150 to 300; under strong benefit, it increased from $0.619$ to $0.920$. When the sign differed by hierarchy, the selected-rule analysis rejected with probabilities $0.197$ and $0.374$ even though the minimum population effect was $-0.120$. The proposed test correctly gave essentially no evidence for a conclusion that held under every rule. The scenario with different signs by threshold showed the same distinction. The minimum point estimator had visible downward bias when several population effects were equal or close, as predicted by Theorem~\ref{thm:minimum}, but the confidence procedure remained conservative in those settings.

\begin{table}[t]
\centering
\caption{Simulation results. Coverage refers to a one-sided $97.5\%$ lower confidence bound. Rejection probabilities use one-sided $\alpha=0.025$. Bias is multiplied by 100.}
\label{tab:simulation}
\resizebox{\textwidth}{!}{%
\begin{tabular}{lrrrrrr}
\toprule
Scenario & $n$ per arm & $\theta_{\min}$ & Bias $\times100$ & Coverage & Proposed test & Selected rule \\
\midrule
Global null & 150 & 0.000 & -2.35 & 0.992 & 0.008 & 0.027 \\
Global null & 300 & 0.000 & -1.60 & 0.994 & 0.006 & 0.026 \\
One active null rule & 150 & -0.002 & -0.28 & 0.974 & 0.025 & 0.227 \\
One active null rule & 300 & -0.002 & -0.05 & 0.974 & 0.023 & 0.408 \\
Weak benefit under all rules & 150 & 0.065 & -1.81 & 0.990 & 0.097 & 0.215 \\
Weak benefit under all rules & 300 & 0.065 & -1.28 & 0.991 & 0.193 & 0.376 \\
Strong benefit under all rules & 150 & 0.151 & -0.94 & 0.985 & 0.619 & 0.729 \\
Strong benefit under all rules & 300 & 0.151 & -0.51 & 0.985 & 0.920 & 0.952 \\
Different signs by hierarchy & 150 & -0.120 & -0.06 & 0.978 & 0.000 & 0.197 \\
Different signs by hierarchy & 300 & -0.120 & 0.04 & 0.977 & 0.000 & 0.374 \\
Different signs by threshold & 150 & -0.051 & -0.55 & 0.983 & 0.002 & 0.021 \\
Different signs by threshold & 300 & -0.051 & -0.17 & 0.983 & 0.001 & 0.022 \\
\bottomrule
\end{tabular}%
}
\end{table}

\subsection{Continuous-threshold example}

The main simulation used three clinically selected thresholds. That design is adequate when those three values constitute the complete prespecified set. A finite list and a full interval define different analyses, as illustrated by the following population example.

Treatment reduced the log odds of death by 0.50 and the log mean hospitalization count by 1.00. The quality-of-life treatment shift was 0.40 for 50\% of participants, $-0.90$ for 35\%, and 2.00 for 15\%; independent residual standard deviations were 0.10 in both groups. This produces small common improvements, moderate less common deterioration, and large improvements in a minority. As the threshold increases, these parts of the quality-of-life distribution enter and leave the pairwise comparison at different points. The resulting net-benefit curve need not be monotone.

We approximated the population curve with five million independent treatment and control pairs and evaluated 101 equally spaced thresholds over $[0.25,1.25]$. The minimum over the three-point grid $\{0.25,0.75,1.25\}$ was 0.016 at 0.75, whereas the interval minimum was $-0.011$ near 0.65. All inspected grid points were favorable, but the declared interval included thresholds with an unfavorable effect.

\begin{figure}[t]
\centering
\includegraphics[width=0.72\textwidth]{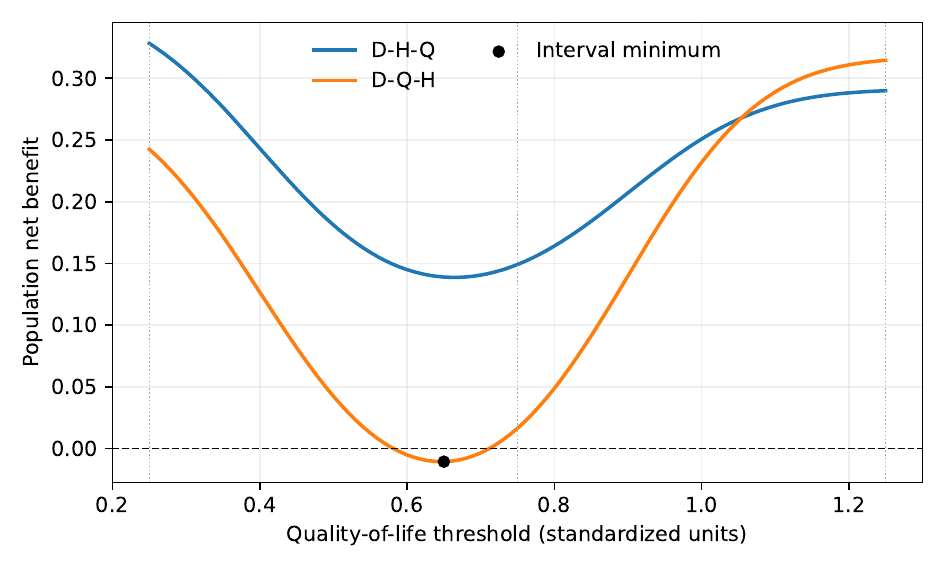}
\caption{Population net benefit over the prespecified quality-of-life threshold interval. Vertical dotted lines show the three values in the sparse grid; the black point marks the interval minimum.}
\label{fig:interval-curve}
\end{figure}

Figure~\ref{fig:interval-curve} also shows the limitation of checking only the interval endpoints. The effect can have an interior minimum because changing a threshold can remove favorable and unfavorable pairwise decisions at different rates. The clinical statement determines the statistical rule set: a finite list calls for a finite analysis, whereas a genuine interval calls for interval-wide evaluation.

\section{Numerical illustration}

We generated one trial with 250 participants per group under the scenario in which the sign differed by hierarchy. The illustration uses simulated data and the same six prespecified hierarchies and thresholds as the simulation study.

Table~\ref{tab:illustration} shows favorable results when hospitalization preceded quality of life, with estimated win ratios from $1.177$ to $1.306$. When quality of life preceded hospitalization, the lowest-threshold analysis produced more losses than wins and a win ratio of $0.831$. The estimated minimum net benefit was $-0.087$, and its one-sided $97.5\%$ lower confidence bound was $-0.188$. These data therefore failed to support benefit under every prespecified hierarchy and threshold, although some individual analyses were favorable.

\begin{table}[t]
\centering
\caption{Results under each hierarchy and threshold in the numerical illustration. The last four columns give the percentage of treated-control pairs decided at each component or left tied.}
\label{tab:illustration}
\resizebox{\textwidth}{!}{%
\begin{tabular}{lrrrrrrr}
\toprule
Order & $\tau_Q$ & Net benefit & Win ratio & Death & Hosp. & Quality of life & Tie \\
\midrule
D--H--Q & 0.25 & 0.077 & 1.177 & 23.6\% & 32.0\% & 39.1\% & 5.3\% \\
D--H--Q & 0.75 & 0.087 & 1.230 & 23.6\% & 32.0\% & 28.8\% & 15.7\% \\
D--H--Q & 1.25 & 0.100 & 1.306 & 23.6\% & 32.0\% & 19.8\% & 24.6\% \\
D--Q--H & 0.25 & -0.087 & 0.831 & 23.6\% & 4.0\% & 67.1\% & 5.3\% \\
D--Q--H & 0.75 & -0.039 & 0.912 & 23.6\% & 11.7\% & 49.1\% & 15.7\% \\
D--Q--H & 1.25 & 0.012 & 1.033 & 23.6\% & 18.4\% & 33.4\% & 24.6\% \\
\bottomrule
\end{tabular}%
}
\end{table}

The percentages in Table~\ref{tab:illustration} explain the reversal. Death decided 23.6\% of pairs under every hierarchy because it was always first. When hospitalization was second, it decided 32.0\% of pairs. When quality of life was second, it decided 67.1\% of pairs at the lowest threshold, leaving only 4.0\% to hospitalization. The hierarchy therefore changes which component decides a large fraction of pairs.

\begin{figure}[t]
\centering
\includegraphics[width=0.58\textwidth]{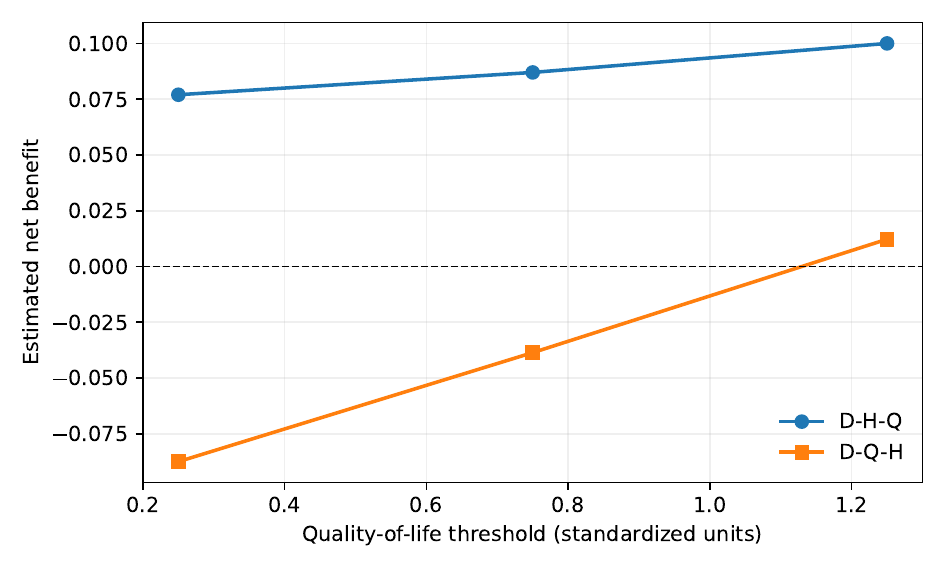}
\caption{Estimated net benefit by hierarchy and quality-of-life threshold in the numerical illustration. The horizontal line marks no net benefit.}
\label{fig:preference-curve}
\end{figure}

Figure~\ref{fig:preference-curve} displays the same information as a function of the quality-of-life threshold. The persistent separation between the two curves indicates a strong contribution from outcome ordering to the conclusion. A clinical analysis should accompany such a plot with the component-level proportions of pairs decided at each tier, because these proportions explain why a lower outcome has substantial influence.

\section{ACTG 175 illustration}

ACTG 175 was a randomized, double-blind trial of four antiretroviral regimens in adults with human immunodeficiency virus infection \citep{hammer1996actg}. We used the public dataset distributed by the UCI Machine Learning Repository \citep{uci2023actg175}. Among the randomized groups, the comparison of zidovudine plus zalcitabine with zidovudine monotherapy provides a compact example in which two laboratory outcomes point in different directions. The groups contained 524 and 532 participants, respectively. Changes from baseline in CD4 and CD8 cell counts at week 20 were recorded for every participant in these groups. This was an exploratory, post hoc illustration.

Two outcome orders were evaluated: CD4 followed by CD8 and CD8 followed by CD4. Larger increases were favorable. A common threshold of 25, 50, or 75 cells per cubic millimeter was applied to both outcomes, giving six analyses. The thresholds were illustrative; a confirmatory trial would need clinical justification for each value. Mean CD4 change was 19.3 cells per cubic millimeter with zidovudine plus zalcitabine and $-17.1$ with zidovudine. The corresponding mean CD8 changes were $-82.5$ and $-59.0$.

Table~\ref{tab:actg175} shows a clear effect of outcome order. With CD4 first, net benefits ranged from 0.091 to 0.156 and all three lower confidence limits were positive. All three CD8-first estimates were negative. Across the six analyses, the estimated minimum net benefit was $-0.039$, its one-sided $97.5\%$ lower confidence bound was $-0.108$, and the intersection-union p-value was 0.863. The negative CD8-first results determined the assessment across the complete rule set.

\begin{table}[t]
\centering
\caption{ACTG 175 results comparing zidovudine plus zalcitabine with zidovudine monotherapy. The threshold, expressed in cells per cubic millimeter, was applied to both laboratory outcomes. Lower limits are one-sided $97.5\%$ confidence limits.}
\label{tab:actg175}
\begin{tabular}{lrrrr}
\toprule
Order & Threshold & Net benefit & Win ratio & Lower limit \\
\midrule
CD4--CD8 & 25 & 0.156 & 1.374 & 0.088 \\
CD4--CD8 & 50 & 0.123 & 1.292 & 0.055 \\
CD4--CD8 & 75 & 0.091 & 1.218 & 0.024 \\
CD8--CD4 & 25 & $-0.039$ & 0.925 & $-0.108$ \\
CD8--CD4 & 50 & $-0.028$ & 0.945 & $-0.097$ \\
CD8--CD4 & 75 & $-0.019$ & 0.960 & $-0.088$ \\
\bottomrule
\end{tabular}
\end{table}

\section{Discussion}

Win-statistic results can depend on the order of the outcomes and on the thresholds used to define wins and losses. This creates a practical problem when more than one comparison rule is clinically reasonable. A favorable result under one rule may become weaker or reverse under another. We address this problem by prespecifying the acceptable rules and taking the smallest net benefit among them. A positive lower confidence bound then shows that treatment is favored under every planned rule. The calculation uses the usual pairwise win and loss scores and the covariance among the rule-specific U-statistics.

In the simulations, the one-sided lower confidence bound had at least the nominal coverage. It was somewhat conservative when several rules had nearly the same smallest effect and was close to the usual one-rule analysis when one rule was clearly least favorable. In ACTG 175, the CD4-first rules favored treatment, whereas the CD8-first rules favored control. The minimum net benefit therefore did not support a favorable result across the complete rule set. The continuous-threshold example also showed that a sparse grid can miss the least favorable threshold. These results support prespecifying the clinically reasonable rules and reporting their individual results together with the minimum.

\bibliographystyle{apalike}
\bibliography{references}

@article{finkelstein1999,
  title={Combining mortality and longitudinal measures in clinical trials},
  author={Finkelstein, Dianne M and Schoenfeld, David A},
  journal={Statistics in medicine},
  volume={18},
  number={11},
  pages={1341--1354},
  year={1999},
  publisher={Wiley Online Library}
}

@article{pocock2012,
  title={The win ratio: a new approach to the analysis of composite endpoints in clinical trials based on clinical priorities},
  author={Pocock, Stuart J and Ariti, Cono A and Collier, Timothy J and Wang, Duolao},
  journal={European heart journal},
  volume={33},
  number={2},
  pages={176--182},
  year={2012},
  publisher={Oxford University Press}
}

@article{buyse2010,
  title={Generalized pairwise comparisons of prioritized outcomes in the two-sample problem},
  author={Buyse, Marc},
  journal={Statistics in medicine},
  volume={29},
  number={30},
  pages={3245--3257},
  year={2010},
  publisher={Wiley Online Library}
}

@article{luo2015,
  title={An alternative approach to confidence interval estimation for the win ratio statistic},
  author={Luo, Xiaodong and Tian, Hong and Mohanty, Surya and Tsai, Wei Yann},
  journal={Biometrics},
  volume={71},
  number={1},
  pages={139--145},
  year={2015},
  publisher={Oxford University Press}
}

@article{bebu2016,
  title={Large sample inference for a win ratio analysis of a composite outcome based on prioritized components},
  author={Bebu, Ionut and Lachin, John M},
  journal={Biostatistics},
  volume={17},
  number={1},
  pages={178--187},
  year={2016},
  publisher={Oxford University Press}
}

@article{oakes2016,
  title={On the win-ratio statistic in clinical trials with multiple types of event},
  author={Oakes, David},
  journal={Biometrika},
  volume={103},
  number={3},
  pages={742--745},
  year={2016},
  publisher={Oxford University Press}
}

@article{mao2019,
  title={On the alternative hypotheses for the win ratio},
  author={Mao, Lu},
  journal={Biometrics},
  volume={75},
  number={1},
  pages={347--351},
  year={2019},
  publisher={Wiley Online Library}
}

@article{dong2023,
  title={Win statistics (win ratio, win odds, and net benefit) can complement one another to show the strength of the treatment effect on time-to-event outcomes},
  author={Dong, Gaohong and Huang, Bo and Verbeeck, Johan and Cui, Ying and Song, James and Gamalo-Siebers, Margaret and Wang, Duolao and Hoaglin, David C and Seifu, Yodit and M{\"u}tze, Tobias and others},
  journal={Pharmaceutical Statistics},
  volume={22},
  number={1},
  pages={20--33},
  year={2023},
  publisher={Wiley Online Library}
}

@article{verbeeck2023,
  title={Generalized pairwise comparisons to assess treatment effects: JACC review topic of the week},
  author={Verbeeck, Johan and De Backer, Mickael and Verwerft, Jan and Salvaggio, Samuel and Valgimigli, Marco and Vranckx, Pascal and Buyse, Marc and Brunner, Edgar},
  journal={Journal of the American College of Cardiology},
  volume={82},
  number={13},
  pages={1360--1372},
  year={2023},
  publisher={American College of Cardiology Foundation Washington DC}
}

@article{mao2024estimand,
  title={Defining estimand for the win ratio: Separate the true effect from censoring},
  author={Mao, Lu},
  journal={Clinical Trials},
  volume={21},
  number={5},
  pages={584--594},
  year={2024},
  publisher={SAGE Publications Sage UK: London, England}
}

@article{shoji2025,
  title={Win ratio analyses using a modified hierarchical composite outcome: insights from PARAGLIDE-HF},
  author={Shoji, Satoshi and Cyr, Derek D and Hernandez, Adrian F and Velazquez, Eric J and Ward, Jonathan H and Williamson, Kristin M and Sarwat, Samiha and Starling, Randall C and Desai, Akshay S and Zieroth, Shelley and others},
  journal={American Heart Journal},
  volume={280},
  pages={70--78},
  year={2025},
  publisher={Elsevier}
}

@article{mao2026poset,
  title={Win ratio for partially ordered data},
  author={Mao, Lu},
  journal={Statistica Sinica},
  volume={37},
  number={1},
  year={2023}
}

@article{mou2025rotation,
  title={Integrating Prioritized and Non-Prioritized Structures in Win Statistics},
  author={Mou, Yunhan and Hummel, Scott and Huang, Yuan},
  journal={arXiv preprint arXiv:2512.18946},
  year={2025}
}

@article{mou2026thresholds,
  title={Generalizing the Finkelstein--Schoenfeld Test to Incorporate Multiple Alternating Thresholds},
  author={Mou, Yunhan and Kyriakides, Tassos and Hummel, Scott and Li, Fan and Huang, Yuan},
  journal={Biometrical Journal},
  volume={68},
  number={2},
  pages={e70117},
  year={2026},
  publisher={Wiley Online Library}
}

@article{ozenne2025,
  title={Overview of the package BuyseTest},
  author={Ozenne, Brice},
  year={2019}
}

@article{arcones1993,
  title={Limit theorems for U-processes},
  author={Arcones, Miguel A and Gin{\'e}, Evarist},
  journal={The Annals of Probability},
  pages={1494--1542},
  year={1993},
  publisher={JSTOR}
}

@article{fang2019,
  title={Inference on directionally differentiable functions},
  author={Fang, Zheng and Santos, Andres},
  journal={The Review of Economic Studies},
  volume={86},
  number={1},
  pages={377--412},
  year={2019},
  publisher={Oxford University Press}
}

@article{chernozhukov2013,
  title={Intersection bounds: Estimation and inference},
  author={Chernozhukov, Victor and Lee, Sokbae and Rosen, Adam M},
  journal={Econometrica},
  volume={81},
  number={2},
  pages={667--737},
  year={2013},
  publisher={Wiley Online Library}
}

@article{berger1982,
  title={Multiparameter hypothesis testing and acceptance sampling},
  author={Berger, Roger L},
  journal={Technometrics},
  volume={24},
  number={4},
  pages={295--300},
  year={1982},
  publisher={Taylor \& Francis}
}

@incollection{vdvaartwellner1996,
  title={Weak convergence},
  author={Van Der Vaart, Aad W and Wellner, Jon A},
  booktitle={Weak convergence and empirical processes: with applications to statistics},
  pages={16--28},
  year={1996},
  publisher={Springer}
}

@article{ehrenzeller2026,
  title={Randomized Clinical Trials Using a Hierarchical Composite Primary End Point: A Scoping Review},
  author={Ehrenzeller, Selina and de Jong, Amos J and Martin, Yonas and de Kraker, Marlieke EA and Jackson, Holly and Hassoun-Kheir, Nasreen and Schumacher, Michaela and Bernasconi, Nadine Saskia and Hensgens, Marjolein PM and Appenzeller-Herzog, Christian and others},
  journal={JAMA Network Open},
  volume={9},
  number={7},
  pages={e2625935},
  year={2026},
  publisher={American Medical Association}
}

@article{hammer1996actg,
  author  = {Hammer, Scott M. and Katzenstein, David A. and Hughes, Michael D. and Gundacker, Holly and Schooley, Robert T. and Haubrich, Richard H. and Henry, W. Keith and Lederman, Michael M. and Phair, John P. and Niu, Margaret and Hirsch, Martin S. and Merigan, Thomas C.},
  title   = {A trial comparing nucleoside monotherapy with combination therapy in {HIV}-infected adults with {CD4} cell counts from 200 to 500 per cubic millimeter},
  journal = {New England Journal of Medicine},
  year    = {1996},
  volume  = {335},
  number  = {15},
  pages   = {1081--1090},
  doi     = {10.1056/NEJM199610103351501}
}

@misc{uci2023actg175,
  author       = {{AIDS Clinical Trials Group Study 175 Study Team}},
  title        = {{AIDS Clinical Trials Group Study 175} [Dataset]},
  year         = {2023},
  howpublished = {UCI Machine Learning Repository},
  doi          = {10.24432/C5ZG8F}
}

\section*{Appendix S1: Proofs}

Equation and theorem numbers refer to the main article.

\subsection*{Proof of Theorem 1}

For a fixed rule $r$, the two-sample Hoeffding decomposition gives
\begin{equation}
\widehat\theta_r-\theta_r
=\frac{1}{n_1}\sum_{i=1}^{n_1}a_r(X_i)
+\frac{1}{n_0}\sum_{j=1}^{n_0}b_r(Y_j)
+R_{nr},
\label{eq:hoeffding}
\end{equation}
where $R_{nr}$ is the degenerate two-sample remainder. Because $|h_r|\leq1$ and both sample sizes are proportional to $n$,
\[
E(R_{nr}^2)=O\{(n_1n_0)^{-1}\},
\qquad
\sqrt nR_{nr}=o_p(1).
\]
Since the number of rules is fixed, the result holds jointly over all rules. The multivariate central limit theorem applied separately to the treatment and control projections gives
\[
\sqrt n(\widehat{\boldsymbol\theta}-\boldsymbol\theta)
=\frac{\sqrt n}{n_1}\sum_{i=1}^{n_1}\boldsymbol a(X_i)
+\frac{\sqrt n}{n_0}\sum_{j=1}^{n_0}\boldsymbol b(Y_j)
+o_p(1),
\]
with the covariance stated in the main article. For the covariance estimator, the row and column averages converge in mean square to the corresponding first-order projections. The two sample covariance matrices therefore converge to $\operatorname{Var}\{\boldsymbol a(X)\}$ and $\operatorname{Var}\{\boldsymbol b(Y)\}$. Multiplication by $n/n_1$ and $n/n_0$ gives the stated consistency result.

\subsection*{Proof of Theorem 2}

Let $\phi(u)=\min_{1\leq r\leq M}u_r$. For any fixed vector $z$,
\begin{equation}
\lim_{t\downarrow0}
\frac{\phi(\boldsymbol\theta+tz)-\phi(\boldsymbol\theta)}{t}
=\min_{r\in\Rzero}z_r.
\label{eq:directional-derivative}
\end{equation}
Thus $\phi$ is Hadamard directionally differentiable at $\boldsymbol\theta$. The directional delta method applied to Theorem 1 gives the stated limit. If $\Rzero$ contains one rule, the limit is one component of the Gaussian vector and is normal. If $\Rzero$ contains more than one rule, the limit is the minimum of the corresponding correlated normal components.

\subsection*{Proof of Theorem 3}

The two-sample Hoeffding decomposition holds uniformly over $r\in\Rset$:
\begin{equation}
\widehat\theta_r-\theta_r
=\frac{1}{n_1}\sum_{i=1}^{n_1}a_r(X_i)
+\frac{1}{n_0}\sum_{j=1}^{n_0}b_r(Y_j)
+R_n(r).
\label{eq:uniform-hoeffding}
\end{equation}
Boundedness and the assumed asymptotic equicontinuity imply
\begin{equation}
\sup_{r\in\Rset}\sqrt n|R_n(r)|=o_p(1).
\label{eq:uniform-remainder}
\end{equation}
The empirical-process central limit theorem applied to the two independent projection classes gives weak convergence of the first two terms in $\ell^\infty(\Rset)$. Their covariance kernels add, with factors $\lambda^{-1}$ and $(1-\lambda)^{-1}$. The same Glivenko--Cantelli conditions and the uniform remainder give
\[
\sup_{r\in\Rset}|\widehat\theta_r-\theta_r|\to_p0.
\]

Because $\Rset$ is compact and $r\mapsto\theta_r$ is continuous, $\Rzero$ is nonempty and compact. Define $\phi(f)=\inf_{r\in\Rset}f(r)$ on $\ell^\infty(\Rset)$. At a continuous function $\theta$, its directional derivative along a continuous direction $g$ is
\begin{equation}
\phi'_{\theta}(g)=\inf_{r\in\Rzero}g(r).
\label{eq:inf-derivative}
\end{equation}
To verify this derivative, any sequence of approximate minimizers of $\theta+t g$ has a convergent subsequence whose limit lies in $\Rzero$, which gives one inequality. Evaluating $\theta+t g$ at a minimizer of $\theta$ gives the reverse inequality. The directional functional delta method then proves the stated limit.

\subsection*{Proof of Theorem 4}

Fix any $r_0\in\Rzero$. By construction,
\[
L_{1-\alpha}
\leq \widehat\theta_{r_0}-z_{1-\alpha}\widehat s_{r_0}.
\]
The rule-specific U-statistic central limit theorem and variance consistency give
\[
P\left\{
\widehat\theta_{r_0}-z_{1-\alpha}\widehat s_{r_0}
\leq\theta_{r_0}
\right\}\longrightarrow1-\alpha.
\]
Because $\theta_{r_0}=\thetamin$, the coverage result follows. Under the null, choose a rule $r_0$ with $\theta_{r_0}\leq0$. The event $L_{1-\alpha}>0$ requires the rule-specific lower limit for $r_0$ to exceed zero, whose limiting probability is no greater than $\alpha$. In the finite-rule case, if exactly one rule has $\theta_r=0$ and all others are separated from zero by a positive constant, the other lower limits exceed zero with probability tending to one, and the rejection probability converges to $\alpha$. When several rules are on the boundary, all corresponding lower limits must exceed zero, so the limiting rejection probability can be smaller.

\end{document}